\documentclass[structabstract]{aa}  
\usepackage{graphicx}
\usepackage{txfonts}
\begin{document}

%\title{Soft X-ray variability of accreting black holes in the 
%High and Very High State}

\title{Stabilization of radiation pressure dominated accretion disks by viscous fluctuations}

 \author{Agnieszka Janiuk\inst{1}
          \and
Ranjeev Misra\inst{2}
}

%affiliation

   \institute{Center for Theoretical Physics, Polish Academy of Sciences, Al. Lotnikow 32/46, 02-668 Warsaw, Poland \\
\email{agnes@cft.edu.pl}
\and
 Inter-University Centre for Astronomy and Astrophysics, Pune 41107, India\\
\email{rmisra@iucaa.ernet.in}
}

   \date{Received ...; accepted ...}
  \abstract
{}
{The
standard thin accretion disk model has been successfully used to explain the
soft X-ray spectra of Galactic black hole systems and perhaps the UV emission of
Active Galactic Nuclei. However, radiation pressure dominated  disks are known to
be viscously unstable and should produce large amplitude oscillations that are
typically not observed. Instead, these sources exhibit stochastic variability which 
may naturally arise due to  viscous fluctuations in a turbulent disk. Here we investigate
whether these aperiodic viscous fluctuations can stabilize the inner radiation pressure
dominated disks and hence maybe the answer to a forty year old problem in accretion disk
theory.
}
   {The structure and evolution
 of a time dependent accretion disk around a black hole 
is solved numerically. 
We incorporate fluctuations in the disk by 
considering stochastic variations in the viscous parameter $\alpha$ on the
local viscous time-scale. We study both locally stable disks where the viscous
stress scales with the gas pressure and the locally unstable disks where the stress scales
as the total pressure. We consider steady state disk parameters pertaining to both
stellar and super-massive black holes. 
}
{
For locally stable disks, the power spectra of the luminosity variations are found to 
inversely scale with frequency i.e. $P(f) \propto 1/f$ as expected from analytical studies of
linear viscous fluctuations. For unstable disks, where the viscous stress scales with
the total pressure, the standard oscillatory solutions are seen when the viscous fluctuation
amplitude is small. Increasing the fluctuation amplitude decreases the amplitude of these
oscillations till for a sufficiently large (near unity) fluctuation amplitude, the oscillatory
behavior disappears and the  luminosity variation of the disk becomes 
stochastic.   
This study may explain
 why many accreting black holes, 
despite having accretion rates large enough to develop the 
radiation pressure instability, do not exhibit large amplitude regular 
outbursts, or 
like the micro-quasar GRS 1915+105, produce diminished versions of them only 
in some 
particular spectral states.
}
{}
  \keywords{X-ray Binaries --
    Galaxies:active --
    black holes --
    accretion disks
  }
  \authorrunning{A. Janiuk \& R. Misra}
  \titlerunning{Variability of accreting black holes}
   \maketitle

\section{Introduction}

The standard accretion disk model (Shakura \& Sunyaev 1973) 
has been the basic framework by which 
the accretion process around black holes is understood. The model, which is also
referred to as the $\alpha-$disk model, assumes a thin Keplerian accretion disk where
the unknown turbulent viscosity is parametrized such that the viscous torque is
proportional to the total pressure i.e. $T_{\rm r \phi} = \alpha P_{\rm tot}$. The model
predicts that for super-massive black holes the peak radiation will be emitted in
the ultra-violet and this is often identified with the  "Big Blue Bump" in Active Galactic Nuclei (AGN).
For stellar mass black holes, the emission peaks in soft X-rays and such a thermal soft
emission is indeed observed in Galactic black hole binaries. 

Galactic black hole binaries exhibit different spectral states  
(see e.g., for a review, Done et al. 2007). 
In the hard/low state the spectra is dominated
by an hard X-ray power-law that extends to $\sim 100$ keV and  a weaker soft black body like  component. 
The soft component is generally modeled as a truncated standard accretion disk and the primary
hard power-law is believed to arise from a hot inner disk 
(e.g., Shapiro, Lightman \& Eardley 1976).
In the soft
state, the spectrum is dominated by a thermal emission from the standard accretion disk. For X-ray novae,
luminosity changes by several orders of magnitude, yet the inferred inner radius of the standard
accretion disk remains constant, as expected from the standard disk theory. 
In fact, 
spectral computation based on the standard theory, but incorporating general relativistic effects and detailed radiative transfer have
been used successfully used to fit the soft component.
Such analysis can constrain the mass and spin of the black holes and opens up
the possibility of testing General Relativity in the strong field limit. Thus, while there may be some ambiguity regarding the
origin of the hard X-ray emission from black hole binaries, there is near consensus that the soft component is due to a standard
accretion disk, which allows for a detailed and unprecedented understanding of these sources.

There has been a long standing concern regarding the standard accretion disk model, that the disk is both
thermally and viscously unstable when radiation pressure is dominant (Lightman \& Eardley 1974). 
This occurs in the inner
regions of the disk around a ten solar mass black hole, when its luminosity exceeds $ \gtrsim 3 \times 10^{37}$ ergs/s.
X-ray binaries can typically exceed this luminosity which is substantially smaller than than the Eddington limit
of $\sim 10^{39}$ ergs/s. For AGNs, harboring $10^7 M_\odot$ black holes, the disk should be unstable for
luminosities exceeding  $\gtrsim 5 \times 10^{43}$ ergs/s significantly less than the Eddington limit of
$\sim 10^{45}$ ergs/s. Thus, for typical black hole binaries and AGNs, the standard accretion disk is unstable
and not hence not applicable.

The extension of the radiation pressure dominated part of the disk
depends on the accretion rate, the black hole mass and the
 viscosity. The parameter regions of the radiation pressure dominated disks 
in Galactic black holes and AGN were shown recently in Janiuk \& Czerny (2011); the 
approximate fitting formula for this dependence is:
\begin{equation}
\log {R_{\rm out} \over R_{\rm S}}=\log{\dot m} +2.5 +0.1 \log{\alpha \over 0.01}
\end{equation}
for a Galactic black hole of 10 $M_{\odot}$ at accretion rates above $\log \dot m_{\rm crit}=-1.5$ where $\dot m = \dot M/\dot M_{Edd}$ and $R_{\rm S}=2 GM/c^{2}$. For a supermassive black hole of $10^{8} M_{\odot}$ at accretion rates above $\log \dot m_{\rm crit}=-2.5$, the radial range is:
\begin{equation}
\log {R_{\rm out} \over R_{\rm S}}=0.5 \log{\dot m} +2.7 +0.1 \log{\alpha \over 0.01}
\end{equation}

Initially it was postulated that the unstable inner regions of the standard disk may get converted
into a hot two temperature region (Shapiro, Lightman \& Eardley 1976). 
The hot inner disk could then be the source of the
hard X-ray emission and is a consistent model for Galactic black holes in the hard state. However, in the
soft state, where the luminosity could be even higher than the hard one, there is a soft thermal-like emission
indicating that the standard disk somehow survives the instability caused by the radiation pressure instability.
Moreover, time-dependent numerical evolution of the unstable disk, shows that instead of the disk getting transformed
to a hot one, it undergoes cyclic behavior, which result in large amplitude oscillations in luminosity. 
These
 large variations in luminosity have not been observed.

Attempts have been made to stabilize the disk, or at least reduce the destabilizing effect, by accounting
for the fraction of power that maybe transferred to an overlying corona or jet. This indeed
increases the critical accretion rate, above which the instability occurs by a factor of few, nevertheless
the instability persists and the model predicts large amplitude oscillations at high luminosities. Further, 
assuming that there is a vertical height dependence on the energy dissipated in the disk, where a significant
fraction of the energy is dissipated near the photosphere and not the mid-plane, also does not completely
stabilize the disk. Finally, one can bring to question, the form of the viscous stress chosen in
the standard model, where the stress is taken to be  proportional to the total pressure, $T_{r\phi} =\alpha P_{\rm tot}$. Indeed, if the
viscous stress scales with only the gas pressure $T_{r\phi} =\alpha P_{\rm gas}$, the disk would be unconditionally stable. However,
it is not clear why such an ad hoc prescription would be a physical one. If the turbulence  length scales with the height of
the disk, but the eddy speed scales with the gas sound speed, $c_{s} = \sqrt{\gamma P_{\rm gas}/\rho}$
 (instead of the total sound speed $c_{s} = \sqrt{\gamma P_{\rm tot}/\rho}$), the viscous stress would be 
$\propto \sqrt{P_{\rm gas}P_{\rm tot}}$. However, even for such a prescription the disk is not stable, but undergoes 
luminosity oscillations with diminished amplitude. A combination of such a viscous prescription and assuming that
a fraction of the power is transferred to a corona, may explain the variability of the micro-quasar GRS 1915+105,
in certain spectral states (Janiuk \& Czerny 2011). GRS 1915+105 is not 
typical and the apparent stability of other bright X-ray binaries
and even for GRS 1915+105 in its stable states, remains a puzzle (see e.g, Neilsen et al. 2011).

Instead of showing large amplitude oscillatory behavior, black hole binaries and AGNs exhibit stochastic
variability with a high frequency cutoff (Nowak et al. 1999).
In the hard state of the X-ray sources, the photons from the disk
 are being upscattered 
due to the Comptonization in the hot plasma. The dispersion of the 
scattering times may lead to the attenuation of the observed variability 
amplitude. Also, the   dependence of the power density spectrum (PDS) 
with energy  may be due to direct fluctuations in the Comptonizing medium (Nowak et al. 1999).
In the soft state, the observed radiation is mainly from the accretion disk, whose dynamical variability may be coupled to the emission processes in a 
non-linear way. Also, the effect of reflection of the hard X-rays from the disk
surface may lead to a feedback effect and influence the variability amplitude 
(Uttley \& McHardy 2005). For some sources, gravitational light bending
may also be affecting the variability of the source 
(NGC-6-30-15; see Miniutti \& Fabian 2004).
In any case, the amplitude of the observed variability in both Galactic black 
holes and AGN is of the order of $30\%$.

One of the models for the physical origin of the disk 
variability is stochastically driven  damped
harmonic oscillations at the local epicyclic frequency (Misra \& Zdziarski 2008).

Alternatively or additionally, the variability may
be due to variations in the local viscous stress, which induces  local accretion rate fluctuations, which in turn
induces and adds to  variability downstream at smaller radii (Lyubarskii  1997). The  model predicts a $1/f$ like variability for
the inner disk accretion rate and a cutoff at the frequency corresponding to the  viscous time-scale of the
inner region. Indeed, the observed power spectra of black hole binaries and AGNs can be explained in terms
of this viscous fluctuation models (e.g., Churazov et al.2001; Uttley et al.2005; Titarchuk et. al. 2007, Ingram \& Done 2011). 
More interestingly, since in this model, the disturbance travels
from the outer to the inner regions and if higher energy photons are preferentially produced in the
inner regions, there would be time-lag between the high and low energy photons. The cylindrical geometry
would make this time-lag inversely proportional to frequency (Misra 2000). The observed time-lags in black hole binaries
and AGNs has this frequency dependence (Nowak et al. 1999) which gives strong credibility to the viscous fluctuation model,
which otherwise would be difficult to explain. Theoretically too, one expects that a turbulent viscous stress, perhaps
arising from the magneto-hydrodynamic rotational instability, should show high amplitude variability as indicated
by numerical simulations. Magneto-Hydrodynamic simulations of 
radiation pressure dominated accretion disks use however 
the shearing box geometry, 
and hence they do not provide information about the nature of the viscous 
instability (Hirose et. al. 2009).

It is interesting to study what effect the viscous stress fluctuations will have on the overall stability of
radiation pressure dominated disks. If the viscous fluctuation amplitude is small, a linear analysis would
naturally show that the unstable mode and viscous fluctuations would be uncoupled and the variability
will simply be their sum. In other words, in the linear regime, the perturbations would grow due to
the unstable mode and would not be affected by the viscous fluctuations. However, when the viscous 
fluctuation amplitudes are large, there may be non-trivial and non-linear coupling between the two,
perhaps leading to interesting results. This may especially be true when the systems evolves due to the
instability and the  viscous time-scale on which the fluctuations occur changes in the disk. Thus, for
a proper understanding of these effects, we undertake here a numerical computation to simulate the 
time-dependent behavior of a standard accretion disk with viscous fluctuations.

The article is organized as follows. In Section 2. we describe our model 
for the viscosity fluctuations, giving also the references to a series of 
previous work where our radiation pressure instability numerical model was
presented and developed.
In Section 3., we present the results, showing first (Sec. 3.1.) the case of 
stable accretion disk models with viscosity fluctuations, and second (Sec. 3.2)
the case of unstable models with fluctuations. In Section 4, we discuss the
results and their implications.

\section{The Numerical Scheme}

We use the 1(+1)-D hydrodynamical code,
developed and described first 
by Janiuk et al. (2002) for the  time-dependent evolution of an accretion disk. 
Depending on the mean accretion rate, the disk may become unstable due to the
radiation pressure domination.
The code has been generalized to incorporate two phase flows taking into account
mass exchange between the disk and an overlying corona (Janiuk \& Czerny 2005)
 and the possibility
of a fraction of the power being transmitted to the a jet/outflow 
(e.g. Kunert-Bajraszewska et al. 2010)
As in the standard model, the code assumes that the flow is Keplerian but allows
for the possibility of energy advection to smaller radii. 

We adopt here the kinematic viscosity of Shakura \& Sunyaev (1973), as less or equal to the sound speed times the disk vertical scaleheight, 
via the dimensionless proportionality parameter $\alpha$.
We incorporate viscous fluctuations in the disk by considering a time-varying 
$\alpha$ parameter given by (see Lyubarskii 1997):

\begin{equation}
\alpha(r,t) = \alpha_{0} [1+\beta(r,t)]
\label{eq:alphart}
\end{equation}
where $\alpha_{0}$ is a standard constant value on the order of 0.01 - 0.1, 
and the small factor $\beta(r,t)$ accounts for the fluctuating part.
whose 
time variation is assumed to follow the Markov chain process:
\begin{equation}
\beta_{n} = b_{0} u_{n}
\label{eq:beta}
\end{equation} 
where $b_{0}$ is a parameter and 
\begin{equation}
u_{n} = -0.5 u_{n-1}+ \epsilon_{n}.
\end{equation}
The factor 0.5 in the above equation ensures that the Markov chain does not diverge for large $n$. In general, the modulus of this 
factor should be less than 1.0, and its value governs the number of 
series steps between the peaks, $N$, as well as the amplitude of the 
oscillations, $var u$. For factor 0.5, we have $N \approx 2.6$ and $var u = 0.75 var \epsilon$ (see e.g. King et al. 2004 and Janiuk \& Czerny 2007).
The random variable $\epsilon_{\rm n}$ is generated with a uniform 
distribution between -1 and 1.
The timescale of the fluctuation is given by the local viscous time-scale at that radius $r$ expressed in terms
of the Schwarzschild radius $2GM/c^2$, as:
\begin{equation}
\tau_{\rm visc} = {1 \over \alpha \Omega} ({r \over H})^{2} f(r)
\end{equation}
Here $H$ is the height, $\Omega$ the Keplerian angular velocity and  
$f(r)=1-\sqrt{3/r}$ is the
 boundary condition to assure that the torque vanishes at
the last stable orbit, which also implies that the density in the disk
goes to zero at its inner edge. 
The coherent length over which $\alpha$ varies is assumed to be the
local height $H$, which is calculated from the total pressure as
$H=c_{s}/\Omega$.

The computations are numerically expensive and were performed with the message passing interface technique, 
on multi-CPU Linux computer cluster machines. We had to restrict the number of radius bins to
be $\sim 200$. For typical parameters used in this work, this translates to a bin size $\Delta r \sim H_o (r)/3$,
where $H_o (r)$ is the height of the steady state disk. This radial grid is kept fixed through the time-evolution
i.e. we do not adaptively change the radial grid.   However, in the model, $\alpha$ variability is taken to be coherent
over  a region $\sim H$, where $H$ is the height at a given time. We take this into account by incorporating  
a multiplicative factor on $\beta$ of $\sqrt{H_0/H}$, which is typically small. This factor increases/decreases the viscous variability 
when the coherent length scale is smaller/larger than the grid size bin.  This scheme is similar to one used by 
Mayer \& Pringle (2006) to model magnetic dynamo evolution.

\section{Results}

\subsection{Stable accretion disk models with viscosity fluctuations}

We first study the effect of viscous fluctuations on stable accretion disks. 
Ideally,
one should analyze accretion disks where the accretion is low, such that the 
disk is
not radiation pressure dominated. However, for such disks the height $H << r$ 
and hence
the numerical computation would require a large number of radial bins, to 
correctly incorporate
the viscous fluctuations. Hence, we consider disks where the viscous stress 
scales only
with gas pressure i.e. $T_{r\phi} = \alpha P_{\rm gas}$. 

\begin{figure}
\includegraphics[width=7cm]{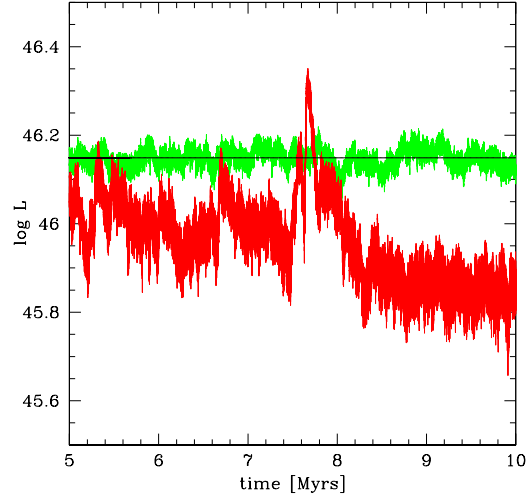}
\caption{Lightcurves of models for stable disks where the viscous stress scales
only with the gas pressure.
The green (red) line shows the model with viscous fluctuation amplitude
 $b_{0}=0.6$ (0.99), and the thick black line shows the model with $b_{0}=0.0$.
The mean viscosity parameter $\alpha_{0}=0.1$,
the mean  accretion rate is $\dot m=0.33$ and the black hole mass is $3\times 10^{8} M_{\odot}$.
}
\label{fig:lum3}
\end{figure}

\begin{figure}
\includegraphics[width=7cm]{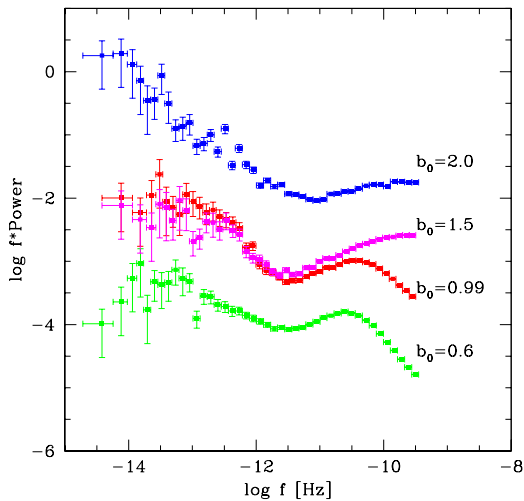}
\caption{Power density spectra of the luminosity fluctuations
for  stable disks where the viscous stress scales
only with the gas pressure.
The models have flickering parameters of $b_{0}=0.6$, 
 $b_{0}=0.99$,
 $b_{0}=1.5$ and $b_{0}=2.0$, as shown with green, red, magenta, and blue 
points, respectively.
The mean viscosity parameter $\alpha_{0}=0.1$,
the mean  accretion rate is $\dot m=0.33$ and the black hole mass is $3\times 10^{8} M_{\odot}$.
}
\label{fig:pds}
\end{figure}

\begin{figure}
   \centering
\includegraphics[width=7cm]{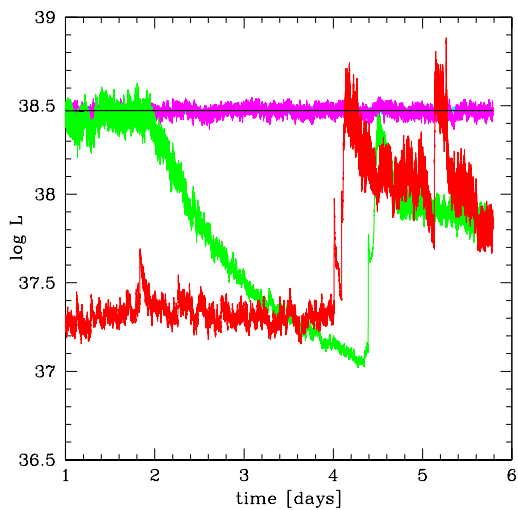}
\caption{Lightcurves of models for stable disks where the viscous stress scales
only with the gas pressure with flickering parameter
$b_{0}=0.3$, 0.6 and 0.99, marked with magenta, green and red lines, respectively, 
and the thick black line shows the model with $b_{0}=0.0$
The mean viscosity parameter $\alpha_{0}=0.1$,
the mean  accretion rate is $\dot m=0.33$ and the black hole mass is $10 M_{\odot}$.
}
\label{fig:lumgbh1}
\end{figure}

\begin{figure}
\includegraphics[width=7cm]{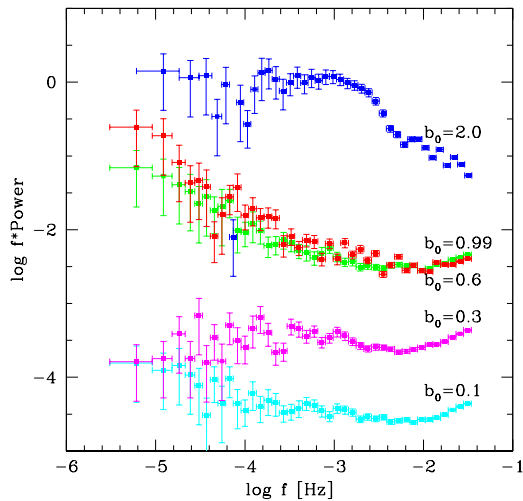}
\caption{Power density spectra of the luminosity fluctuations
 for stable disks where the viscous stress scales
only with the gas pressure.
The models have flickering parameters of $b_{0}=0.1$, 
 $b_{0}=0.3$,  $b_{0}=0.6$,
 $b_{0}=0.99$ and $b_{0}=2.0$, as shown with cyan, magenta, green, red and blue 
points, respectively.
The mean viscosity parameter $\alpha_{0}=0.1$,
the mean  accretion rate is $\dot m=0.33$ and the black hole mass is $10 M_{\odot}$.
}
\label{fig:pdsgbh}
\end{figure}

In the presence of viscous fluctuations the luminosity of the disk varies, as shown in
in Figure \ref{fig:lum3}, where we plot the lightcurves of a disk around a $3 \times 10^8 M_{\odot}$ black hole
with accretion rate $\dot m =  0.33$ in Eddington units and  $\alpha_{0}=0.1$. As expected, as
the flickering amplitude $b_0$ increases so does the variability of the luminosity. It is
more illustrative to see the power density of the luminosity fluctuations which are shown
in  Figure \ref{fig:pds}. For small variations in the viscous fluctuations, the
variation in the luminosity can be estimated to have a power spectrum, roughly $\propto 1/f$ 
(Lyubarskii 1997). This is qualitatively consistent with the power spectra shown in
Figure \ref{fig:pds}, especially for small $b_0$, i.e. the power spectrum is
roughly $\propto 1/f$ such that $f \times$ power is nearly independent of frequency.
For larger values of $b_0$ there is a significant increase of power in  low frequency.
The reason for this behavior is more evident when we plot the light curves
for a disk around a $10 M_{\odot}$ black hole in Figure \ref{fig:lumgbh1} and
the corresponding power spectra in Figure \ref{fig:pdsgbh}. The lightcurves show that
for large $b_0$ apart from increased stochastic variability, there are also long time
scale ``jumps'' in the luminosity. These low frequency variations which occur in the
viscous time-scale of the outer radii, may indeed be due to outer boundary conditions imposed
in the numerical scheme. The outer disk radius for the computations has been taken
to be $\sim 1000$ Schwarzschild radii. Increasing the outer radii, incorporates a large
cost in computation time especially since the total simulation time has to be
increased to be longer than the viscous time-scale of the outer radius.

While the low frequency variability at large $b_0$ may be an artifact of
the outer boundary conditions, we note that it will not effect the results of our
primary study, which is the effect of fluctuations on unstable disks. Here the instability
time-scale is the viscous time-scale of the inner accretion disk region which is significantly
shorter than the viscous time-scale of the outer radius. 

Thus, in general, our numerically computed luminosity variations of  the viscous fluctuations 
are consistent with that expected from analytical estimates that predict the power spectra
$\propto 1/f$.

\subsection{Unstable accretion disk models with viscosity fluctuations}

\begin{figure}
   \centering
\includegraphics[width=9cm]{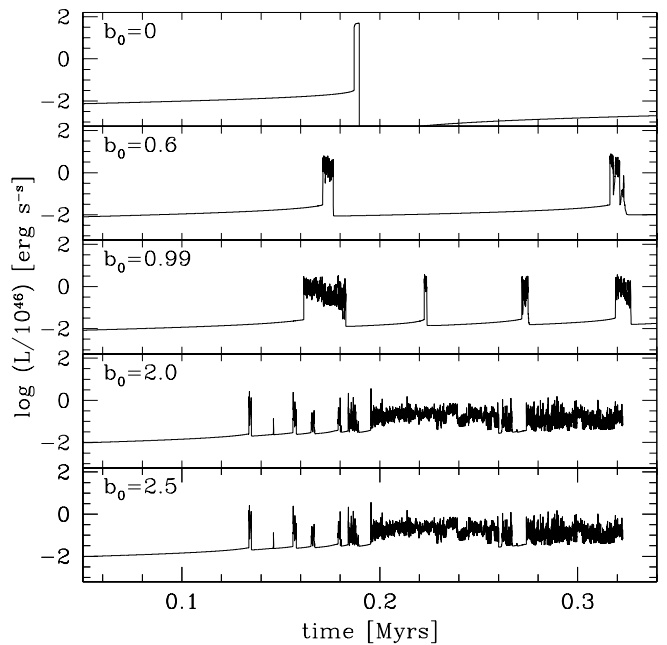}
\caption{Lightcurves in the 5 models of accretion disk
oscillations due to the radiation pressure instability limit-cycle.
The results are shown for
the alpha flickering models with parameters $b_{0}=0.0$, $b_{0}=0.6$, $b_{0}=0.99$, $b_{0}=2$ and $b_{0}=2.5$, from top to bottom,
respectively (the quiescent period between thew outbursts 
in $b_{0}=0.0$ model is about 0.6 Myrs).
The model parameters are: black hole mass of $3\times 10^{8} M_{\odot}$, accretion rate of $\dot m = 0.33$
in Eddington units, $\alpha_{0}=0.1$. The model includes no jet.}
\label{fig:lummqagn}
\end{figure}

Having verified that the results of our numerical code are generally consistent with that
expected for stable accretion disk where analytical results are possible, we consider
the more physically consistent situation where the viscous stress scales with the
total pressure. Our primary result is illustrated in Figure \ref{fig:lummqagn}.
 For low values of viscous fluctuation amplitude $b_0 = 0.66$, the
disk is unstable and shows periodic very large amplitude outbursts as expected.

It can be seen from the plot that the strong fluctuations in the
luminsoity occurs when the flux level of the source is high. This is
because during such ``high'' flux states, the disk is thicker and
hence the stochastic viscous fluctuations occur over a larger
coherent range. Moreover, the viscous time-scale is shorter and hence
the effects of viscous fluctuations are more pronounced.

Increasing $b_0 = 0.99$  causes the oscillations to become irregular and the
duration of the bursts also changes. Further increase of $b_0 = 2.0$ decreases
 the amplitude of the bursts and the lightcurve becomes more ``chaotic'' in nature.
Finally for a large $b_0 = 2.5$, the lightcurve after a transient beginning phase,
becomes stochastic and there is no evidence for any cyclic bursts.

The time averaged accretion rate in all the simulations shown in Figure 
\ref{fig:lummqagn} is equal to the external accretion rate supplied to the disk
($\dot m=0.33$). When the radiation pressure instability operates, 
the disk luminosity cannot be constant, corresponding to this average
 acretion rate,
 but switches between the low and high luminosities, corresponding to the 
stages when the accretion rate onto the black hole is either much higher 
or much lower than the average value. When the instability is suppressed by the viscous fluctuations, there are no longer any 'low' or 'high' states.

\begin{figure}
   \centering
\includegraphics[width=9cm]{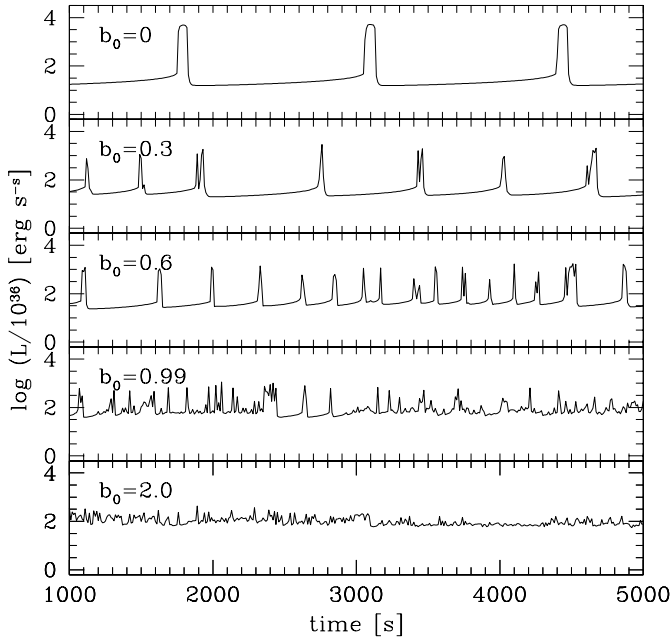}
\caption{Lightcurves in the 5 models of accretion disk
oscillations due to the radiation pressure instability limit-cycle.
The results are shown for
the alpha flickering models with parameters $b_{0}=0.0$, $b_{0}=0.3$, $b_{0}=0.6$, $b_{0}=0.99$ and $b_{0}=2.0$, from top to bottom,
respectively.
The model parameters are: black hole mass of $10 M_{\odot}$, 
accretion rate of $\dot m = 0.33$
in Eddington units, $\alpha_{0}=0.1$. The model includes no jet.}
\label{fig:lummq}
\end{figure}

For stellar mass black holes, we find similar results with the disk showing oscillatory
behavior for low values of $b_0$ and stochastic behavior for higher values
(see Figure \ref{fig:lummq}). The
disks stabilize at a lower value of $b_{0} \approx 1$ as compared to $b_{0} \ge 1.5$ for AGN disks.
This maybe related to the fact that at the same Eddington rate, an AGN disk is
more radiation pressure dominated than that for a stellar mass black hole.

We also consider the effect of power being transferred to a jet.
Specifically, we parametrize the jet (cf. Janiuk \& Czerny 2011)
by including an additional term in the energy equation
\begin{equation}
 Q^{-}={c P_{\rm rad}\over \kappa \Sigma}(1-\eta_{\rm jet})^{-1}
\end{equation}
where the fraction $\eta_{\rm jet}$ is assumed to depend on 
 on time and distance from the black hole 
via the local accretion rate:
\begin{equation}
\eta_{\rm jet} = 1 - {1 \over 1+  A_{\rm jet} \dot m(r,t)^{2}}.
\label{eq:ajet}
\end{equation}
with $A_{\rm jet}$ being a parameter of the model. The physical 
meaning of this parametrization is that at the Eddington accretion limit 
there would be equipartition between the jet and disk radiation for 
$A_{\rm jet}=1$. We adopt a value of $A_{\rm jet}=5$ or $=10$ , 
which means a moderately strong jet. Without viscous fluctuations, even for such
moderately strong jets the disk undergoes cyclic oscillations, but with a reduced
amplitude. The inclusion of comparatively small viscous fluctuations $b_0 = 0.3$ stabilizes
the disk for both values of $A_{jet}$. Thus a moderately strong jet decreases the
amount of viscous fluctuation required for stability. 

In Table \ref{table:1} we present the summary of the models we have 
calculated. In summary, large amplitude viscous fluctuations stabilize the disk
against the radiation pressure instability. The strength of the viscous fluctuation
required to stabilize the disk is slightly lower for stellar mass black hole systems
than AGNs and decreases when there is an effective jet in the system. The
Table also shows the normalized r.m.s variability of the luminosity. Note that
for locally stable disks (i.e. when $\tau_{r\phi} \propto P_{\rm gas}$), the r.m.s increases
with $b_o$, while for locally unstable disks(i.e. when $\tau_{r\phi} \propto P_{\rm tot}$), 
it generally decreases. This is because the viscous fluctuations stabilize the disk
and high amplitude oscillations disappear.

%
%_____________________________________________________________
%                                             Two column Table 
%_____________________________________________________________
%
\begin{table}
\caption{Summary of the time-dependent disk models. The accretion rate in Eddington units is 
$\dot m =0.33$ and mean viscosity is $\alpha_{0}=0.1$ for each case. r.m.s. is the normalized
root mean square variability of the luminosity.  }             
\label{table:1}      
\centering          
\begin{tabular}{c c c c  c c }   
\hline\hline       
Mass [$M_{\odot}$] & $A_{jet}$ & $b_{0}$ & $\tau_{r\phi}\propto$ & r.m.s. & Outbursts\\
\hline    
& & & & & \\                
%-------------------------------------
    $3\times 10^{8}$ & 0 & 0.6&  $P_{gas}$ & 0.05   & no  \\ 
     $3\times 10^{8}$ & 0 & 0.99& $P_{gas}$ & 0.28  & no  \\
     $3\times 10^{8}$ & 0 & 1.5 & $P_{gas}$ & 0.73   & no \\ 
    $3\times 10^{8}$ & 0 & 2.0 & $P_{gas}$& 1.92   & no \\ 
& & & & & \\
%-------------------------------------
    $3\times 10^{8}$ & 0 & 0.6 & $P_{tot}$&  5.0  & yes  \\ 
 $3\times 10^{8}$ & 0 & 0.99& $P_{tot}$ & 2.2 & yes  \\ 
    $3\times 10^{8}$ & 0 & 1.5 & $P_{tot}$& 1.63 & no \\
   $3\times 10^{8}$ & 0 & 2.0 & $P_{tot}$ &1.5  & no \\ 
    $3\times 10^{8}$ & 0 & 2.5 & $P_{tot}$& 0.92   & no \\ 
& & & & & \\
%-----------------------------------------------------------------
    10  &    0           &  0.1  &  $P_{gas}$ & 0.02  & no \\ 
    10  &    0           &  0.3 &  $P_{gas}$ & 0.06   & no \\ 
   10  &    0           &  0.6  &  $P_{gas}$ & 0.79 & no \\ 
    10  &    0           &  0.99  &  $P_{gas}$& 1.13  & no \\ 
    10  &    0           &  2.0  &  $P_{gas}$ & 2.12  & no \\ 
& & & & & \\
%-------------------------------------
    10 &    0          &  0.1 &  $P_{tot}$& 4.82   & yes \\ 
    10 &    0          &  0.3 &  $P_{tot}$& 3.49  & yes \\ 
    10 &    0          &  0.6 &  $P_{tot}$& 2.88   & yes \\ 
    10 &    0          &  0.99 &  $P_{tot}$& 1.72   & no\\ 
    10 &    0          &  2.0  &  $P_{tot}$& 2.81  & no \\ 
& & & & & \\
%-------------------------------------
    10 &    5           &  0.1  &  $P_{tot}$& 1.17 & yes \\  
    10 &    5           &  0.3 &  $P_{tot}$ &0.75  & no \\ 
    10 &    5           &  0.6 &  $P_{tot}$ & 0.55  & no \\ 
    10 &    5           &  0.99 &  $P_{tot}$ & 0.43  & no \\ 
& & & & & \\
%-------------------------------------
    10 &    10           &  0.1  &  $P_{tot}$& 0.94 & yes \\  
    10 &    10           &  0.3 &  $P_{tot}$& 0.72 & no \\ 
    10 &    10           &  0.6 &  $P_{tot}$& 0.52 & no \\ 
%-------------------------------------
\hline                  
\end{tabular}
\end{table}

We defer a detailed comparison of these results with the observations of a 
particular source for future work. This maybe done for the  various states of 
the microquasar GRS 1915+105 (Belloni et al. 2000). Another good candidate
maybe the recently studied source IGR J17091-3624, which also exhibits the 
so-called 'heartbeat' states for luminosity variations in some spectral states,
as well as  
stochastic fluctuations (Altamirano et al. 2011). The detailed modeling would
require a proper determination of the Eddington ratio as well as an estimation
of  the fraction of energy going into jet power from radio observations.

\section{Summary}

Our numerical simulations show that large amplitude viscous fluctuations can stabilize
 radiation pressure dominated standard accretion disks. This may explain why even at
luminosities close to the Eddington, black hole systems exhibit spectral signatures of a standard
accretion disk which do not show large amplitude cyclic behavior. Instead, they show stochastic
variation which maybe due to viscous fluctuations which stabilize the disk.

At low luminosities, the disk would be gas pressure dominated and hence intrinsically stable.
Here, viscous fluctuations may produce variability in the luminosity. However, since it is likely
that the coherent length of the viscous variability scales as the height of the disk which is
much smaller than the radius for low accretion rate systems, the net effect on the global
luminosity should be small. In other words, since $H/R << 1$ for these sources, the viscous
fluctuations at different radii will incoherently add to give a diminishing effect on the
total luminosity of the disk. This may explain why the standard disk component in some low luminosity
systems are found to be steady with little variability.

While, in this work, we consider some general characteristics of Galactic black holes and AGNs,
a more detailed comparison with data of one or two specific sources may provide both
qualitative and quantitative insights. In particular, the time-varying spectra and not just
the integrated luminosity from these
numerical simulations should be compared with data from some chosen sources. The model should
be confronted with both the variability as a function of energy as well as frequency dependent
time-lags.   

In the light of the results of this work, it will also be interesting to re-examine
the models for GRS 1915+105 for cases when the system shows oscillations as well as
when it does not. Detailed comparison between the expected and observed lightcurves,
may reveal whether the viscous variability amplitude changes between the different
states or it is the fraction of energy going into the jet. The study may finally reveal
the distinguishing feature of the source as compared to other black hole systems.

Finally, a more comprehensive numerical analysis needs to be undertaken with different
boundary conditions and assumptions, that may identify the crucial conditions necessary for
the unstable disk to be stabilized by viscous fluctuations. An adaptive radial grid will
be in particular useful. The rotating disk  may also give rise  to non-trivial effects, which
can be studied only by a full 3-D hydrodynamic code. Indeed, what seems to be important is that
the non-linear effect of large amplitude perturbations, which can be captured only
by numerical simulations, can give rise to non-trivial results such as stabilization of
a system which is locally highly unstable.

\begin{acknowledgements}
We thank Bozena Czerny for helpful discussions.
Part of this work was done thanks to the hospitality of the
Inter University Center for Astronomy and Astrophysics, Pune, India.
Partially this work was supported by grant NN 203 512638 from the
Polish National Science Center.
A part of the numerical computations have been 
performed on the computer cluster 
machines in the Copernicus Astronomical Center of PAS in Warsaw.
\end{acknowledgements}

\end{document}